


%
%


\def\famname{
 \textfont0=\textrm \scriptfont0=\scriptrm
 \scriptscriptfont0=\sscriptrm
 \textfont1=\textmi \scriptfont1=\scriptmi
 \scriptscriptfont1=\sscriptmi
 \textfont2=\textsy \scriptfont2=\scriptsy \scriptscriptfont2=\sscriptsy
 \textfont3=\textex \scriptfont3=\textex \scriptscriptfont3=\textex
 \textfont4=\textbf \scriptfont4=\scriptbf \scriptscriptfont4=\sscriptbf
 \skewchar\textmi='177 \skewchar\scriptmi='177
 \skewchar\sscriptmi='177
 \skewchar\textsy='60 \skewchar\scriptsy='60
 \skewchar\sscriptsy='60
 \def\rm{\fam0 \textrm} \def\bf{\fam4 \textbf}}
\def\sca#1{scaled\magstep#1} \def\scah{scaled\magstephalf} 
\def\twelvepoint{
 \font\textrm=cmr12 \font\scriptrm=cmr8 \font\sscriptrm=cmr6
 \font\textmi=cmmi12 \font\scriptmi=cmmi8 \font\sscriptmi=cmmi6 
 \font\textsy=cmsy10 \sca1 \font\scriptsy=cmsy8
 \font\sscriptsy=cmsy6
 \font\textex=cmex10 \sca1
 \font\textbf=cmbx12 \font\scriptbf=cmbx8 \font\sscriptbf=cmbx6
 \font\it=cmti12
 \font\sectfont=cmbx12 \sca1
 \font\sectmath=cmmib10 \sca2
 \font\sectsymb=cmbsy10 \sca2
 \font\refrm=cmr10 \scah \font\refit=cmti10 \scah
 \font\refbf=cmbx10 \scah
 \def\twelverm{\textrm} \def\twelveit{\it} \def\twelvebf{\textbf}
 \famname \textrm 
\advance\voffset by .06in \advance\hoffset by .12in
 \normalbaselineskip=17.5pt plus 1pt \baselineskip=\normalbaselineskip
 \parindent=21pt
 \setbox\strutbox=\hbox{\vrule height10.5pt depth4pt width0pt}}


\catcode`@=11

{\catcode`\'=\active \def'{{}^\bgroup\prim@s}}

\def\screwcount{\alloc@0\count\countdef\insc@unt}   
\def\screwdimen{\alloc@1\dimen\dimendef\insc@unt} 
\def\screwbox{\alloc@4\box\chardef\insc@unt}

\catcode`@=12


\overfullrule=0pt			
\vsize=9in \hsize=6in
\lineskip=0pt				
\abovedisplayskip=1.2em plus.3em minus.9em 
\belowdisplayskip=1.2em plus.3em minus.9em	
\abovedisplayshortskip=0em plus.3em	
\belowdisplayshortskip=.7em plus.3em minus.4em	
\parindent=21pt
\setbox\strutbox=\hbox{\vrule height10.5pt depth4pt width0pt}
\def\makefootline{\baselineskip=30pt \line{\the\footline}}
\footline={\ifnum\count0=1 \hfil \else\hss\twelverm\folio\hss \fi}
\pageno=1


\def\boxit#1{\leavevmode\thinspace\hbox{\vrule\vtop{\vbox{\hrule%
	\vskip3pt\kern1pt\hbox{\vphantom{\bf/}\thinspace\thinspace%
	{\bf#1}\thinspace\thinspace}}\kern1pt\vskip3pt\hrule}\vrule}%
	\thinspace}
\def\Boxit#1{\noindent\vbox{\hrule\hbox{\vrule\kern3pt\vbox{
	\advance\hsize-7pt\vskip-\parskip\kern3pt\bf#1
	\hbox{\vrule height0pt depth\dp\strutbox width0pt}
	\kern3pt}\kern3pt\vrule}\hrule}}


\def\put(#1,#2)#3{\screwdimen\unit  \unit=1in
	\vbox to0pt{\kern-#2\unit\hbox{\kern#1\unit
	\vbox{#3}}\vss}\nointerlineskip}


\def\\{\hfil\break}
\def\newpage{\vfill\eject}
\def\center{\leftskip=0pt plus 1fill \rightskip=\leftskip \parindent=0pt
 \def\textindent##1{\par\hangindent21pt\footrm\noindent\hskip21pt
 \llap{##1\enspace}\ignorespaces}\par}
\def\unnarrower{\leftskip=0pt \rightskip=\leftskip}

%
%
%
%
%
%
%

\def\thetitle#1#2#3#4#5{
 \font\titlefont=cmbx12 \sca2 \font\footrm=cmr10 \font\footit=cmti10
  \twelverm
	{\hbox to\hsize{#4 \hfill YITP-SB-#3}}\par
	\vskip.8in minus.1in {\center\baselineskip=1.44\normalbaselineskip
 {\titlefont #1}\par}{\center\baselineskip=\normalbaselineskip
 \vskip.5in minus.2in #2
	\vskip1.4in minus1.2in {\twelvebf ABSTRACT}\par}
 \vskip.1in\par
 \narrower\par#5\par\unnarrower\vskip3.5in minus2.3in\eject}
\def\paper\par#1\par#2\par#3\par#4\par#5\par{\twelvepoint
	\thetitle{#1}{#2}{#3}{#4}{#5}} 
\def\author#1#2{#1 \vskip.1in {\twelveit #2}\vskip.1in}
\def\YITP{C. N. Yang Institute for Theoretical Physics\\
	State University of New York, Stony Brook, NY 11794-3840}
\def\WS{W. Siegel\footnote{${}^1$}{       
	\pdflink{mailto:siegel@insti.physics.sunysb.edu}\\
	\pdfklink{http://insti.physics.sunysb.edu/\~{}siegel/plan.html}
	{http://insti.physics.sunysb.edu/\noexpand~siegel/plan.html}}}


\def\sect#1\par{\par\ifdim\lastskip<\medskipamount
	\bigskip\medskip\goodbreak\else\nobreak\fi
	\noindent{\sectfont{#1}}\par\nobreak\medskip} 
\def\itemize#1 {\item{[#1]}}	
\def\vol#1 {{\refbf#1} }		 

\def\ref#1{\setbox0=\hbox{M}$\vbox to\ht0{}^{#1}$}


\def\NP #1 {{\refit Nucl. Phys.} {\refbf B{#1}} }
\def\PL #1 {{\refit Phys. Lett.} {\refbf{#1}} }
\def\PR #1 {{\refit Phys. Rev. Lett.} {\refbf{#1}} }
\def\PRD #1 {{\refit Phys. Rev.} {\refbf D{#1}} }


\hyphenation{pre-print}
\hyphenation{quan-ti-za-tion}

%
%


\def\oonoo#1#2#3{\vbox{\ialign{##\crcr
	\hfil\hfil\hfil{$#3{#1}$}\hfil\crcr\noalign{\kern1pt\nointerlineskip}
	$#3{#2}$\crcr}}}
\def\oon#1#2{\mathchoice{\oonoo{#1}{#2}{\displaystyle}}
	{\oonoo{#1}{#2}{\textstyle}}{\oonoo{#1}{#2}{\scriptstyle}}
	{\oonoo{#1}{#2}{\scriptscriptstyle}}}
\def\dt#1{\oon{\hbox{\bf .}}{#1}}  
\def\ddt#1{\oon{\hbox{\bf .\kern-1pt.}}#1}    
\def\slap#1#2{\setbox0=\hbox{$#1{#2}$}
	#2\kern-\wd0{\hfuzz=1pt\hbox to\wd0{\hfil$#1{/}$\hfil}}}
\def\sla#1{\mathpalette\slap{#1}}                
\def\bop#1{\setbox0=\hbox{$#1M$}\mkern1.5mu
	\lower.02\ht0\vbox{\hrule height0pt depth.06\ht0
	\hbox{\vrule width.06\ht0 height.9\ht0 \kern.9\ht0
	\vrule width.06\ht0}\hrule height.06\ht0}\mkern1.5mu}
\def\bo{{\mathpalette\bop{}}}                        
\def~{\widetilde} 
\mathcode`\*="702A                  
\def\in{\relax\ifmmode\mathchar"3232\else{\refit in\/}\fi} 
\def\f#1#2{{\textstyle{#1\over#2}}}	   
\def\half{{\textstyle{1\over{\raise.1ex\hbox{$\scriptstyle{2}$}}}}}

\def\Gamma{\mathchar"0100}
\def\Delta{\mathchar"0101}
\def\Theta{\mathchar"0102}
\def\Lambda{\mathchar"0103}
\def\Xi{\mathchar"0104}
\def\Pi{\mathchar"0105}
\def\Sigma{\mathchar"0106}
\def\Upsilon{\mathchar"0107}
\def\Phi{\mathchar"0108}
\def\Psi{\mathchar"0109}
\def\Omega{\mathchar"010A}

\catcode`\^^?=13				    
\catcode128=13 \def €{\"A}                 
\catcode129=13 \def {\AA}                 
\catcode130=13 \def '{\c}           	   
\catcode131=13 \def ƒ{\'E}                   
\catcode132=13 \def "{\~N}                   
\catcode133=13 \def …{\"O}                 
\catcode134=13 \def †{\"U}                  
\catcode135=13 \def ‡{\'a}                  
\catcode136=13 \def ˆ{\`a}                   
\catcode137=13 \def ‰{\^a}                 
\catcode138=13 \def Š{\"a}                 
\catcode139=13 \def ‹{\~a}                   
\catcode140=13 \def Œ{\alpha}            
\catcode141=13 \def {\chi}                
\catcode142=13 \def Ž{\'e}                   
\catcode143=13 \def {\`e}                    
\catcode144=13 \def {\^e}                  
\catcode145=13 \def '{\"e}                
\catcode146=13 \def '{\'\i}                 
\catcode147=13 \def "{\`\i}                  
\catcode148=13 \def "{\^\i}                
\catcode149=13 \def •{\"\i}                
\catcode150=13 \def –{\~n}                  
\catcode151=13 \def —{\'o}                 
\catcode152=13 \def ˜{\`o}                  
\catcode153=13 \def ™{\^o}                
\catcode154=13 \def š{\"o}                 
\catcode155=13 \def ›{\~o}                  
\catcode156=13 \def œ{\'u}                  
\catcode157=13 \def {\`u}                  
\catcode158=13 \def ž{\^u}                
\catcode159=13 \def Ÿ{\"u}                
\catcode160=13 \def  {\tau}               
\catcode161=13 \mathchardef ¡="2203     
\catcode162=13 \def ¢{\oplus}           
\catcode163=13 \def £{\relax\ifmmode\to\else\itemize\fi} 
\catcode164=13 \def ¤{\subset}	  
\catcode165=13 \def ¥{\infty}           
\catcode166=13 \def ¦{\mp}                
\catcode167=13 \def §{\sigma}           
\catcode168=13 \def ¨{\rho}               
\catcode169=13 \def ©{\gamma}         
\catcode170=13 \def ª{\leftrightarrow} 
\catcode171=13 \def «{\relax\ifmmode\acute\else\expandafter\'\fi}
\catcode172=13 \def ¬{\relax\ifmmode\expandafter\ddt\else\expandafter\"\fi}
\catcode173=13 \def ­{\equiv}            
\catcode174=13 \def ®{\approx}          
\catcode175=13 \def ¯{\Omega}          
\catcode176=13 \def °{\otimes}          
\catcode177=13 \def ±{\ne}                 
\catcode178=13 \def ²{\le}                   
\catcode179=13 \def ³{\ge}                  
\catcode180=13 \def ´{\upsilon}          
\catcode181=13 \def µ{\mu}                
\catcode182=13 \def ¶{\delta}             
\catcode183=13 \def ·{\epsilon}          
\catcode184=13 \def ¸{\Pi}                  
\catcode185=13 \def ¹{\pi}                  
\catcode186=13 \def º{\beta}               
\catcode187=13 \def »{\partial}           
\catcode188=13 \def ¼{\nobreak\ }       
\catcode189=13 \def ½{\zeta}               
\catcode190=13 \def ¾{\sim}                 
\catcode191=13 \def ¿{\omega}           
\catcode192=13 \def À{\dt}                     
\catcode193=13 \def Á{\gets}                
\catcode194=13 \def Â{\lambda}           
\catcode195=13 \def Ã{\nu}                   
\catcode196=13 \def Ä{\phi}                  
\catcode197=13 \def Å{\xi}                     
\catcode198=13 \def Æ{\psi}                  
\catcode199=13 \def Ç{\int}                    
\catcode200=13 \def È{\oint}                 
\catcode201=13 \def É{\relax\ifmmode\cdot\else\vol\fi}    
\catcode202=13 \def Ê{\relax\ifmmode\,\else\thinspace\fi}
\catcode203=13 \def Ë{\`A}                      
\catcode204=13 \def Ì{\~A}                      
\catcode205=13 \def Í{\~O}                      
\catcode206=13 \def Î{\Theta}              
\catcode207=13 \def Ï{\theta}               
\catcode208=13 \def Ð{\relax\ifmmode\bar\else\expandafter\=\fi}
\catcode209=13 \def Ñ{\overline}             
\catcode210=13 \def Ò{\langle}               
\catcode211=13 \def Ó{\relax\ifmmode\{\else\ital\fi}      
\catcode212=13 \def Ô{\rangle}               
\catcode213=13 \def Õ{\}}                        
\catcode214=13 \def Ö{\sla}                      
\catcode215=13 \def ×{\relax\ifmmode\check\else\expandafter\v\fi}
\catcode216=13 \def Ø{\"y}                     
\catcode217=13 \def Ù{\"Y}  		    
\catcode218=13 \def Ú{\Leftarrow}       
\catcode219=13 \def Û{\Leftrightarrow}       
\catcode220=13 \def Ü{\relax\ifmmode\Rightarrow\else\sect\fi}
\catcode221=13 \def Ý{\sum}                  
\catcode222=13 \def Þ{\prod}                 
\catcode223=13 \def ß{\widehat}              
\catcode224=13 \def à{\pm}                     
\catcode225=13 \def á{\nabla}                
\catcode226=13 \def â{\quad}                 
\catcode227=13 \def ã{\in}               	
\catcode228=13 \def ä{\star}      	      
\catcode229=13 \def å{\sqrt}                   
\catcode230=13 \def æ{\^E}			
\catcode231=13 \def ç{\Upsilon}              
\catcode232=13 \def è{\"E}    	   	 
\catcode233=13 \def é{\`E}               	  
\catcode234=13 \def ê{\Sigma}                
\catcode235=13 \def ë{\Delta}                 
\catcode236=13 \def ì{\Phi}                     
\catcode237=13 \def í{\`I}        		   
\catcode238=13 \def î{\iota}        	     
\catcode239=13 \def ï{\Psi}                     
\catcode240=13 \def ð{\times}                  
\catcode241=13 \def ñ{\Lambda}             
\catcode242=13 \def ò{\cdots}                
\catcode243=13 \def ó{\^U}			
\catcode244=13 \def ô{\`U}    	              
\catcode245=13 \def õ{\bo}                       
\catcode246=13 \def ö{\relax\ifmmode\hat\else\expandafter\^\fi}
\catcode247=13 \def÷{\relax\ifmmode\tilde\else\expandafter\~\fi}
\catcode248=13 \def ø{\ll}                         
\catcode249=13 \def ù{\gg}                       
\catcode250=13 \def ú{\eta}                      
\catcode251=13 \def û{\kappa}                  
\catcode252=13 \def ü{\half}     		 
\catcode253=13 \def ý{\Gamma} 		
\catcode254=13 \def þ{\Xi}   			
\catcode255=13 \def ÿ{\relax\ifmmode{}^{\dagger}{}\else\dag\fi}


\def\ital#1Õ{{\it#1\/}}	     
\def\un#1{\relax\ifmmode\underline#1\else $\underline{\hbox{#1}}$
	\relax\fi}

\def\tdt#1{\oon{\hbox{\bf .\kern-1pt.\kern-1pt.}}#1}   
\def\({\eqno(}

\def\refs{\sect{REFERENCES}\par\medskip \frenchspacing 
	\parskip=0pt \refrm \baselineskip=1.23em plus 1pt
	\def\ital##1Õ{{\refit##1\/}}}


\def\õ#1{
	\screwcount\num
	\num=1
	\screwdimen\downsy
	\downsy=-1.5ex
	\mkern-3.5mu
	õ
	\loop
	\ifnum\num<#1
	\llap{\raise\num\downsy\hbox{$õ$}}
	\advance\num by1
	\repeat}
\def\upõ#1#2{\screwcount\numup
	\numup=#1
	\advance\numup by-1
	\screwdimen\upsy
	\upsy=.75ex
	\mkern3.5mu
	\raise\numup\upsy\hbox{$#2$}}



\newcount\marknumber	\marknumber=1
\newcount\countdp \newcount\countwd \newcount\countht 

%
%
\ifx\pdfoutput\undefined
\def\rgboo#1{}
\input epsf

\def\postscript#1{\special{" #1}}		
\postscript{
	/bd {bind def} bind def
	/fsd {findfont exch scalefont def} bd
	/sms {setfont moveto show} bd
	/ms {moveto show} bd
	/pdfmark where		
	{pop} {userdict /pdfmark /cleartomark load put} ifelse
	[ /PageMode /UseOutlines		
	/DOCVIEW pdfmark}
\def\bookmark#1#2{\postscript{		
	[ /Dest /MyDest\the\marknumber /View [ /XYZ null null null ] /DEST pdfmark
	[ /Title (#2) /Count #1 /Dest /MyDest\the\marknumber /OUT pdfmark}%
	\advance\marknumber by1}
\def\pdfklink#1#2{%
	\hskip-.25em\setbox0=\hbox{#1}%
		\countdp=\dp0 \countwd=\wd0 \countht=\ht0%
		\divide\countdp by65536 \divide\countwd by65536%
			\divide\countht by65536%
		\advance\countdp by1 \advance\countwd by1%
			\advance\countht by1%
		\def\linkdp{\the\countdp} \def\linkwd{\the\countwd}%
			\def\linkht{\the\countht}%
	\postscript{
		[ /Rect [ -1.5 -\linkdp.0 0\linkwd.0 0\linkht.5 ] 
		/Border [ 0 0 0 ]
		/Action << /Subtype /URI /URI (#2) >>
		/Subtype /Link
		/ANN pdfmark}{\rgb{1 0 0}{#1}}}
%
%
\else
\def\rgboo#1{\pdfliteral{#1 rg #1 RG}}

\pdfcatalog{/PageMode /UseOutlines}		
\def\bookmark#1#2{
	\pdfdest num \marknumber xyz
	\pdfoutline goto num \marknumber count #1 {#2}
	\advance\marknumber by1}
\def\pdfklink#1#2{%
	\noindent\pdfstartlink user
		{/Subtype /Link
		/Border [ 0 0 0 ]
		/A << /S /URI /URI (#2) >>}{\rgb{1 0 0}{#1}}%
	\pdfendlink}
\fi

\def\rgbo#1#2{\rgboo{#1}#2\rgboo{0 0 0}}
\def\rgb#1#2{\mark{#1}\rgbo{#1}{#2}\mark{0 0 0}}
\def\pdflink#1{\pdfklink{#1}{#1}}
\def\xxxlink#1{\pdfklink{#1}{http://arXiv.org/abs/#1}}

\catcode`@=11

\def\wlog#1{}	


\def\makeheadline{\vbox to\z@{\vskip-36.5\p@
	\line{\vbox to8.5\p@{}\the\headline%
	\ifnum\pageno=\z@\rgboo{0 0 0}\else\rgboo{\topmark}\fi%
	}\vss}\nointerlineskip}
\headline={
	\ifnum\pageno=\z@
		\hfil
	\else
		\ifnum\pageno<\z@
			\ifodd\pageno
				\tenrm\romannumeral-\pageno\hfil\lefthead\hfil
			\else
				\tenrm\hfil\righthead\hfil\romannumeral-\pageno
			\fi
		\else
			\ifodd\pageno
				\tenrm\hfil\righthead\hfil\number\pageno
			\else
				\tenrm\number\pageno\hfil\lefthead\hfil
			\fi
		\fi
	\fi}

\catcode`@=12

\def\righthead{\hfil} \def\lefthead{\hfil}
\nopagenumbers


\def\chrulefill{\rgb{1 0 0}{\hrulefill}}
\def\cdotfill{\rgb{1 0 0}{\dotfill}}
\newcount\area	\area=1
\newcount\cross	\cross=1
\def\volume#1\par{\newpage\noindent{\biggest{\rgb{1 .5 0}{#1}}}
	\par\nobreak\bigskip\medskip\area=0}
\def\chapskip{\par\ifnum\area=0\bigskip\medskip\goodbreak
	\else\newpage\fi}
\def\chapy#1{\area=1\cross=0
	\xdef\lefthead{\rgbo{1 0 .5}{#1}}\vbox{\biggerer\offinterlineskip
	\line{\chrulefill¼\hphantom{\lefthead}\chrulefill}
	\line{\chrulefill¼\lefthead\chrulefill}}\par\nobreak\medskip}
\def\chap#1\par{\chapskip\bookmark3{#1}\chapy{#1}}
\def\sectskip{\par\ifnum\cross=0\bigskip\medskip\goodbreak
	\else\newpage\fi}
\def\secty#1{\cross=1
	\xdef\righthead{\rgbo{1 0 1}{#1}}\vbox{\bigger\offinterlineskip
	\line{\cdotfill¼\hphantom{\righthead}\cdotfill}
	\line{\cdotfill¼\righthead\cdotfill}}\par\nobreak\medskip}
\def\sect#1 #2\par{\sectskip\bookmark{#1}{#2}\secty{#2}}
\def\subsectskip{\par\ifdim\lastskip<\medskipamount
	\bigskip\medskip\goodbreak\else\nobreak\fi}
\def\subsecty#1{\noindent{\sectfont{\rgbo{.5 0 1}{#1}}}\par\nobreak\medskip}
\def\subsect#1\par{\subsectskip\bookmark0{#1}\subsecty{#1}}
\long\def\x#1 #2\par{\hangindent2\parindent\rgb{0 0 1}{{\bf Exercise #1}\\{#2}}\par}
\def\refs{\bigskip\noindent{\bf \rgbo{0 .5 1}{REFERENCES}}\par\nobreak\medskip
	\frenchspacing \parskip=0pt \refrm \baselineskip=1.23em plus 1pt
	\def\ital##1Õ{{\refit##1\/}}}
\long\def\twocolumn#1#2{\hbox to\hsize{\vtop{\hsize=2.9in#1}
	\hfil\vtop{\hsize=2.9in #2}}}


\twelvepoint
\font\bigger=cmbx12 \sca2
\font\biggerer=cmb10 \sca5
\font\biggest=cmssdc10 scaled 3200

 \sca3


\def Ü{\relax\ifmmode\Rightarrow\else\expandafter\subsect\fi}
\def Û{\relax\ifmmode\Leftrightarrow\else\expandafter\sect\fi}
\def Ú{\relax\ifmmode\Leftarrow\else\expandafter\chap\fi}

\def\itemize#1 {\item{\bf#1}}
\def\itemizze#1 {\itemitem{\bf#1}}
\def\itemutem{\par\indent\indent \hangindent3\parindent \textindent}
\def\itemizzze#1 {\itemutem{\bf#1}}
\def ª{\relax\ifmmode\leftrightarrow\else\itemizze\fi}
\def Á{\relax\ifmmode\gets\else\itemizzze\fi}

\def\¢{\ominus}

\def\Ä{\varphi}  \def\¿{\varpi}

\def ò{\relax\ifmmode\cdots\else\dotfill\fi}




\def\today{\ifcase\month\or
 January\or February\or March\or April\or May\or June\or July\or
 August\or September\or October\or November\or December\fi
 \space\number\day, \number\year}

\parindent=20pt
\newskip\normalparskip	\normalparskip=.7\medskipamount
\parskip=\normalparskip	



\catcode`\|=\active \catcode`\<=\active \catcode`\>=\active 
\def|{\relax\ifmmode\delimiter"026A30C \else$\mathchar"026A$\fi}
\def<{\relax\ifmmode\mathchar"313C \else$\mathchar"313C$\fi}
\def>{\relax\ifmmode\mathchar"313E \else$\mathchar"313E$\fi}



\pageno=0

\paper

\biggest{\rgb{0 0.8 0.5}{Untwisting the twistor superstring}}

\author\WS\YITP

04-20

April 30, 2004

The twistor superstring, which describes N=4 super Yang-Mills trees, is taken off-shell (for loops) by generalizing Penrose twistors (which describe on-shell momenta) to Atiyah-Drinfel'd-Hitchin-Manin twistors (which include the usual spacetime coordinates).  The resulting string is then shown to be the tensionless limit of a Quantum ChromoDynamics-like superstring.

\pageno=2

Ü1. Introduction

Strings were invented for hadrons.  As defined by Dolen-Horn-Schmid duality, strings naturally describe bound states as a consequence of their Regge behavior, and so might describe hadrons as bound states of quarks and gluons without using the quark and gluon fields themselves explicitly.  Unfortunately, no calculable string theory was found to serve that purpose, so string theory got sidetracked into describing fundamental particles, like quarks and gluons.  Apparently, in this picture hadronic strings are thus bound states of partonic strings, which are in turn bound states of preons, ... .  

Alternatively, one might interpret the graviton on the same level as hadrons.  In fact, all observed particles (at least in Grand Unified Theories) can be considered as bound states:  electroweak states already at the classical level, by binding with Higgs fields (for invariance with respect to electroweak gauge invariances); strong states by confinement; and the (super)graviton to preserve renormalizability.  Such a possibility is hinted at by the Anti-de Sitter/Conformal Field Theory correspondence.

A related matter is the scarcity of strings that are naturally (or ``critically") four dimensional.  This contradicts expectations from Quantum ChromoDynamics, since confinement has critical dimension 4, and thus so should hadronic strings.  Of the calculable strings, the only four-dimensional ones are those that are not really strings at all:  strings with local N=2 [1] or 4 [2] worldsheet supersymmetry, and twistor superstrings [3-6].  The latter are not dual, while the former maintain duality only trivially, having vanishing S-matrices [7].

An action for a bosonic string with dimension 4 was proposed [8] based on the use of random lattice quantization of the worldsheet [9], corresponding to bound states in wrong-sign $Ä^4$ theory.  Although the relation of this string to the underlying Feynman diagrams was clear, the method of quantization of the string itself wasn't; i.e., confinement was not demonstrated (which was perhaps just as well for a bosonic theory).  The main distinction of this ``QCD-like" string from other string theories is that it exhibits power behavior (instead of exponential) at large transverse momenta by construction, since its partons have normal propagators instead of the Gaussian ones in other strings.

Twistor strings originated with Nair [3], who proposed a 2D conformal field theory approach to the twistor expressions (as originally found in the spinor helicity formalism [10]) for Maximally Helicity Violating diagrams in Yang-Mills theory (and QCD) [11], extending the results to a supertwistor version for N=4 super Yang-Mills theory.  Meanwhile it was found that the selfdual theories described by N=2 worldsheet supersymmetric strings (and the equivalent N=4 theories) required maximal spacetime supersymmetry to solve various other problems [12].  Since the spacecone gauge formulation of (super) Yang-Mills theory (and QCD) [13], which incorporated spinor helicity methods directly into the action (by reducing fields to ``scalars" of particular helicities), explicitly included separate selfdual and nonselfdual vertices, it suggested a perturbation about selfdual string theories to describe non-selfdual field theories [14].  Such an approach was achieved by Witten [4], who extended Nair's formalism to include non-MHV amplitudes by the use of Dirichlet-branes.  An explicit recipe for the non-MHV amplitudes in this language was then given by Roiban, Spradlin, and Volovich [5].  A simpler formulation was given by Berkovits, who replaced the D-branes with U(1) worldsheet instantons [6].

In this paper we examine twistor superstrings, and solve their two major shortcomings:  (1) They have only (super)twistors as variables, not the usual spacetime coordinates.  Hence they do not seem capable of describing physics off shell.  This makes it doubtful that they can be useful for loop graphs.  (However, lightcone formalisms are similar, in that $p^-$ is eliminated as an independent variable, and it may be possible to apply corresponding methods to twistors.)  Related problems are the description of instantons (also difficult in lightcone formalisms), avoidance of infrared singularities, and a simple description of the fundamental principle of locality.  We will solve these problems by showing the Penrose [15] (super [16])twistors of the twistor superstring can be obtained by solving constraints for and gauge fixing a twistor superstring whose variables are Atiyah-Drinfel'd-Hitchin-Manin [17] (super [18])twistors, the kind used to describe instantons in position space.  (2) They aren't strings.  Although they can be applied to N=4 super (and ordinary) Yang-Mills, it would be nice to have an actual 4D string.  We propose a solution to this problem by showing the ADHM superstring is a tensionless limit [19] of a true superstring, a QCD-like string.  (We do not address the problem of its first-quantization here.)  On the random lattice it yields the Feynman diagrams of a superparticle similar to the one described by the tensionless limit, i.e., N=4 super Yang-Mills in an ADHM formulation.

In the next section we review the twistor superstring, adding some comments.  In the following section we review ADHM supertwistors and apply them to generalize the twistor superstring to a form that includes spacetime coordinates; the usual twistor superstring follows directly from solving constraints (in a manner similar to, but simpler than, lightcone approaches to standard strings).  The next section reviews the random lattice, tensionless limits, and QCD-like strings, and gives the QCD-like superstring whose tensionless limit is the ADHM twistor superstring.  Finally, we summarize and list some avenues of further research.

Ü2. Twistor superstring

The major ingredients in the twistor superstring are: (1) supertwistors as 2D conformal fields, in place of the usual $X$, (2) a current that carries the indices of the Yang-Mills group, but is also responsible for the denominators of MHV amplitudes, and (3) U(1) worldsheet instantons (or the equivalent D-branes), which produce the helicity expansion about the selfdual theory.

One way to understand Penrose twistors is as a covariant solution to the mass-shell condition in the massless case:  In van der Waerden 2-component notation for Weyl spinors,
$$ 0 = p^2 = p^{ŒÀŒ}p_{ŒÀŒ} ¾ det(p^{ŒÀŒ})âÜâp^{ŒÀŒ} = àÂ^Œ ÐÂ^{ÀŒ} $$
where $Â$ is a commuting Weyl spinor (``twistor"), $ÐÂ$ its complex conjugate, and $à$ is for the positive- and negative-energy solutions.  One problem with formalisms where twistors are introduced Óab initioÕ is the loss of this sign; for example, first-quantization can result in a positive-energy propagator rather than the usual St¬uckelberg-Feynman propagator.  However, it might be possible to solve this problem by Wick rotation from spacetime with signature $(--++)$, where $Â$ and $ÐÂ$ are real and independent (SO(2,2)=SL(2,R)${}^2$).  

There is an invariance introduced by this solution, namely
$$ Â' = e^{iÏ}Â,ââÐÂ' = e^{-iÏ}Р$$
This U(1) transformation is a helicity transformation:  Picking a twistor corresponds to choosing a lightlike frame, with helicity the little group.  The phase is replaced by a real scale for the $(--++)$ case.  For the rest of this paper we will mostly ignore such distinctions of Wick rotations; for spacetime symmetries we will use Unitary groups interchangeably with General Linear ones.

Another way to understand twistors, useful especially for extension to supersymmetry, is as the defining representation of the 4D conformal group SU(2,2); i.e., as the spinor representation of SO(4,2).  We quantize by identifying the complex conjugate spinor as also the canonical conjugate,
$$ [Ðñ_A,ñ^BÕ = ¶_A{}^B,ââÐñ_A ­ (ñ^B)ÿú_{ÀBA} $$
where $ú$ is the U(2,2|N) metric, $A=(Œ,ÀŒ,a)$ is a U(2,2|N) index, and we have already generalized to the superconformal case.  The U(2,2|N) generators for this representation are simply
$$ J_A{}^B = ñ^B Ðñ_A $$
from which we factor the superconformal group SU(2,2|N), the remaining U(1) generator being the (super)helicity
$$ H = ñ^A Ðñ_A $$
The previous definition of twistors follows by recognizing momentum as some of these generators.  ($ÐÂ$ is not canonically conjugate to $Â$, because $ú$ is off-diagonal in the basis where Lorentz invariance is manifest.)  Some may be familiar with Schwinger's analogous construction of representations of SU(2) with bosonic creation and annihilation operators [20], the U(1) generator being the spin itself, not its square.  Equivalently, $ñ$ and $Ðñ$ are creation and annihilation operators, and we work with a coherent-state basis.

In Berkovits' formulation the twistor superstring action is
$$ L = (á_- ñ^A)Ðñ_A +L_{YM} $$
where $á_-$ is the ``$-$" (or ``$Ðz$") component of the worldsheet derivative, covariantized with respect to 2D coordinate and U(1) gauge transformations, and $L_{YM}$ is the action for the Yang-Mills symmetry current, which we will not modify, and does not couple to the U(1).  (It can be represented, for example, as a fermionic U(n) current [21].)  The U(1) gauge field thus acts as a Lagrange multiplier to enforce the constraint of the superhelicity to zero; for N=4 supersymmetry, this lowest spin (helicity) multiplet is super Yang-Mills.  Because everything is chiral, there are only massless states.  (The usual massive states of string theory require an $X$ that has a propagator with itself, so that $e^{ikÉX}$ has nontrivial conformal weight.)

Contrary to Berkovits, we interpret this action as a closed-string action; at this point this is semantics:  Berkovits introduced left- and right-handed modes that matched at boundaries; but the boundaries played no role in the theory.  However, the closed-string interpretation will be more natural when we discuss its derivation from an ADHM twistor superstring, where $X$ appears explicitly:  We do not want a separate $X_L$ and $X_R$.  This distinction will be even more pronounced when the ADHM superstring is derived as the tensionless limit of a QCD-like string:  There $X$ appears quadratically, so the usual string interpretation is unavoidable, and such a doubling is undesirable.

Since only left-handed modes exist, vertex operators for this closed string resemble open-string vertex operators:
$$ V(z) = Ä(ñ(z))Éj(z) $$
where $Ä(ñ)$ is the wave function (whose usual twistor argument has been replaced with the 2D twistor field) and $j$ is the Yang-Mills symmetry current; both carry symmetry indices, which we have suppressed, contracted between the two.  We will ignore the contributions of ghosts in this paper.

The U(1) gauge field is analogous to the worldsheet metric:  Although it is formally gauged away in a conformal gauge, it has topological contributions.  Just as the Lorentz connection gauges two symmetries, local Lorentz and local Weyl scale, so the U(1) gauge field gauges left and right U(1)'s with its left and right components.  (For the Lorentz connection, these are scale $à$ Lorentz.)  But the conformal gauge cannot be maintained globally, since the Euler number is gauge invariant, so the worldsheet curvature winds up concentrated at the interaction points or vertex insertions.  For 2D fields with unphysical conformal weight, like worldsheet ghosts, this produces the familiar ghost insertions of string field theory.  For the U(1) gauge field, the Euler number is replaced with worldsheet ``instanton" number [22], and insertions are produced for 2D fields with nonvanishing U(1) charge.  In general, these insertions take the form of $¶$ functions (although for fermions, this is just the fermion itself).  In this case there is an insertion of
$$ ¶^8(Ðñ(z)) $$
for each ÓantiÕ-selfdual vertex, and no insertion for the selfdual ones.  In string field theory $z$ is at the usual midpoint, but because the insertion is BRST invariant (if we had included the ghost contribution) it can be moved about freely on the worldsheet when evaluating S-matrices.  In the conformal field theory calculation it is convenient to put them all at the same point, conventionally $z=¥$.  Because of the equal number of bosons and fermions for N=4, there is no singularity on their coincidence:
$$ ¶^8(Ðñ(z))¶^8(Ðñ(z+·)) = ¶^8(Ðñ(z))¶^8(Ðñ(z)+·Ðñ'(z)) $$
$$ = ¶^8(Ðñ(z))¶^8(·Ðñ'(z)) = ¶^8(Ðñ(z))¶^8(Ðñ'(z)) $$
since the bosons contribute $1/·$'s while the fermions contribute $·$'s.  Such $¶$ insertions also appear in the lightcone superfield theory of N=4 super Yang-Mills, where fermionic insertions for the antiselfdual vertex compensate for the fact that it is naturally written as an integral over $ÐÏ$ rather than $Ï$ [23].  (Similar remarks apply to the lightcone superfield theory of the 10D superstring:  There the vertex insertion is the sum of a term with no fermions, a term with 4 fermions, and a term with 2 fermions that involves momenta in the 6 directions not included in D=4 [24].)

Matrix elements for the $n$-point function are then of the form
$$ A = \leftÒÇ\left[Þ_{i=1}^n{dz_i\over z_i-z_{i+1}}
	Ä_i(ñ(z_i))\right]¶^{8m}(Ðñ(¥))\rightÔ $$
where the $1/(z_i-z_{i+1})$'s are from the Yang-Mills symmetry currents, and the wave functions $Ä_i$ are implicitly cyclically ordered and traced with respect to those symmetry indices.  (There is also a modding out by a GL(2) that consists of the usual SL(2) for the open string and a similar global mode for the U(1).  The $i·$ prescription of the 2D $1/z$ propagators becomes that for the 4D $1/p^2$ internal propagators when the $z$ integration identifies the $z$'s with twistors.)  The U(1) instanton number is $m$, and the product of $m$ insertions produces progressively higher derivatives of $Ðñ$ as above.  The wave functions satisfy the on-shell condition
$$ ñ^A{»\over »ñ^A}Ä_i(ñ) = 0 $$
The conformal weight of $ñ$ in ``flat" 2D space is 0, so that for $Ðñ$ is 1.  This is affected by the U(1) instantons, but this is automatic when the instantons are implemented as insertions.
  
Evaluation of the matrix element involves 2D propagators between $ñ$ and $Ðñ$, and can proceed in either of two ways, but usually as a combination:  For ÓeachÕ SU(2,2|4) component of $ñ$, we expand the wave function in either a $ñ$ basis or a $Ðñ$ basis, by Fourier transformation:
$$ Ä_i(ñ(z_i)) = Çdñ_i¼Ä_i(ñ_i)¶(ñ_i-ñ(z_i))âorâÇdÐñ_i¼÷Ä_i(Ðñ_i)e^{iÐñ_i ñ(z_i)} $$
where $ñ(z_i)$ is the 2D field, and $ñ_i$ and $Ðñ_i$ are dummy variables, arguments of $Ä$'s that are now just wave functions, not vertex operators.  (The Fourier transform form is the standard way vertex operators are evaluated as functions of $X(z,Ðz)$.)  If the Fourier transform form is used, the $¶^{8m}$ insertion is left as is; otherwise, it is written as a Fourier transform
$$ ¶(Ðñ(¥)) = Çd×ñ_1¼e^{-i×ñ_1 Ðñ(¥)} $$
and similarly for derivatives of $Ðñ$.  Evaluation of the propagators is easier if the insertion is at $z=0$, or we can just transform $z£-1/z$ to move it from $¥£0$, evaluate, then transform back.  We then find
$$ \leftÒ\left[ Þ_{i=1}^n ¶(ñ_i-ñ(z_i)) \right] e^{-i×ñ_1 Ðñ(0)} \rightÔ = 
	Þ_{i=1}^n ¶(ñ_i -×ñ_0 -z_i{}^{-1}×ñ_1) $$
where $×ñ_0$ is the zero-mode of $ñ(z)$, which must be integrated over, or
$$ \leftÒ\left[ Þ_{i=1}^n e^{iÐñ_i ñ(z_i)} \right] ¶(Ðñ(0)) \rightÔ = 
	¶\left( Ý_{i=1}^n Ðñ_i \right) ¶\left( Ý_{i=1}^n z_i{}^{-1}Ðñ_i \right) $$
where the first $¶$ comes from integrating over $×ñ_0$.  (The analog for ordinary strings is the usual overall momentum conservation $¶$.)  Both generalize in an obvious way for more insertions:  In the former case $¶(Ðñ')$ is exponentiated with $×ñ_2$, adding a $-z_i{}^{-2}×ñ_2$ to the $¶$'s because of the derivative; in the latter case $¶(Ðñ')$ directly gives a $¶\left( Ýz_i{}^{-2}Ðñ_i \right)$ from differentiating the argument of the second $¶$.  The result for multiple insertions is thus written as either
$$ \leftÒ\left[ Þ_{i=1}^n ¶(ñ_i-ñ(z_i)) \right] ¶^m(Ðñ(0)) \rightÔ = 
	Çd^{m+1} ×ñ_r Þ_{i=1}^n ¶\left( ñ_i -Ý_{r=0}^m z_i{}^{-r}×ñ_r \right) $$
or its Fourier transform
$$ \leftÒ\left[ Þ_{i=1}^n e^{iÐñ_i ñ(z_i)} \right] ¶^m(Ðñ(0)) \rightÔ = 
	Þ_{r=0}^m ¶\left( Ý_{i=1}^n z_i{}^{-r}Ðñ_i \right) $$
(Replacing $z£-1/z$ gives the conventional result.)

To compare with the form of amplitudes obtained by spinor helicity or spacecone methods, one uses the mixed representation $öÄ(Â,ÐÂ)$ in terms of
$$ ñ^A = (Â^Œ,е^{ÀŒ}),ââÐñ_A = (µ_Œ,ÐÂ_{ÀŒ}) $$
(There are also the fermions, which may be mixed or not, depending on one's preference for chiral, antichiral, or twisted chiral superfields.)

Ü3. ADHM twistor superstring

ADHM twistors can also be derived by covariantly solving for the vanishing of the square of a vector.  In this case, it is a 6-vector representing 4D spacetime on the projective lightcone:  Starting with the vector representation $y$ of SO(4,2), we impose the invariant constraint $y^2=0$, and identify $y$'s that differ only by a scale.  If we solve these conditions in the usual lightcone way, we end up with the usual 4 $x$'s, as $x^i=y^i/y^+$.  On the other hand, we can find a spinor solution if we again write the vector in spinor notation:  As an antisymmetric tensor of SU(2,2),
$$ 0 = y^2 = \f14 ·_{ABCD}y^{AB}y^{CD} ¾ å{det(y^{AB})}âÜâ
	y^{AB} = ñ^{Œ'A}ñ_{Œ'}{}^B $$
where $Œ'$ is a new SU(2) index, or SL(2,C) index (contracted with the antisymmetric metric).  The new SU(2) symmetry (which makes U(2) if we include the original scale invariance on $y$), as opposed to the U(1) of Penrose twistors, corresponds to the fact that the little group off-shell is SU(2).  Thus the quaternionic projective space HP(1) describes the usual 4D spacetime, whereas the complex projective space CP(3) describes 3D on-shell momentum space.

As for Penrose twistors, ADHM twistors can also be derived directly as representations of the conformal group, in this case as coset representations.  The basic idea is to to write U(2,2|N) as a square matrix, mod out by two diagonal subgroups, in this case U(1,1) and U(1,1|N), and pick one of the rectangular off-diagonal blocks as creation operators, the twistor.  (It is actually a complex construction, so one uses GL(4|N), with subgroups GL(2) and GL(2|N), and then Wick rotates.)  In analogy to the Penrose case, we have a canonically conjugate $Ðñ_{Œ'A}$,
$$ [Ðñ_{Œ'A},ñ^{º'B}Õ = ¶_{Œ'}^{º'} ¶_A^B $$
but complex conjugation is funny in Minkowski space (but simple for other signatures).  So the superconformal generators are
$$ J_A{}^B = ñ^{Œ'B}Ðñ_{Œ'A} $$
excluding the U(1), which we include in the U(2) constraint
$$ H_{Œ'}{}^{º'} = ñ^{º'A}Ðñ_{Œ'A} $$

The usual way to reproduce $x$ from this twistor is to factor out the U(2) as
$$ ñ_{Œ'}{}^B = U_{Œ'}{}^© (¶_©^º,x_©{}^{Àº},Ï_©{}^b) $$
or choose the U(2) gauge
$$ ñ_{Œ'}{}^º = ¶_{Œ'}^ºâÜâñ_{Œ'}{}^B = (¶_Œ^º,x_Œ{}^{Àº},Ï_Œ{}^b) $$
leaving the coordinates of chiral superspace.  But we can also obtain antichiral superspace by a different factoring/gauge choice:
$$ ñ_{Œ'}{}^{Àº} = ¶_{Œ'}^{Àº}âÜâñ_{Œ'}{}^B = (x^º{}_{ÀŒ},¶_{ÀŒ}^{Àº},ÐÏ_{ÀŒ}{}^b) $$
Alternatively, we can consider the SU(2) invariant, but constrained generalization of the projective lightcone coordinates as above
$$ y^{AB} = ñ^{Œ'A}ñ_{Œ'}{}^B $$
where $A$ is now the full U(2,2|N) index.  In either interpretation, the ADHM twistor contains the complete 4 $x$'s.

Selfdual theories can easily be formulated (at least at the level of equations of motion) in ADHM twistor superspace.  In particular, the ADHM construction for general instanton solutions can be generalized to supersymmetry just by naively extending the $A$ index from bosonic to super.  (Also, in the 't Hooft parametrization for a single instanton, the SU(2) index $Œ'$ can be identified with the Yang-Mills SU(2) index.)  For example, a Yang-Mills connection for the $ñ$ derivative $»_{Œ'A}$ can be introduced, and the selfduality condition can be expressed as
$$ [á_{Œ'A},á_{º'B}Õ = C_{Œ'º'}F_{AB} $$
where $C_{Œ'º'}$ is the SU(2) antisymmetric metric and $F_{AB}$ is the selfdual superfield strength.

Also, the super Klein-Gordon equation is
$$ »^{Œ'}\negthinspace\negthinspace{}_A »_{Œ'B} Ä(ñ) = 0 $$
On fixing the U(2) gauge, these reduce to the usual $õ$, $Ö»d$, and $dd$ equations, in terms of the spinor derivative $d$, which is just $»/»Ï$ in the chiral representation.  These constraints are reducible:  One need solve only the $õ$ and half of the $Ö»d$.

This suggests a simple generalization of the Penrose-twistor superstring to an ADHM-twistor superstring:  The action is
$$ L = (á_- ñ^{Œ'A})Ðñ_{Œ'A} +g^{AB}Ðñ^{Œ'}\negthinspace\negthinspace{}_A Ðñ_{Œ'B} 
	+g_{Œ'}{}^{º'} ñ^{Œ'A}Ðñ_{º'A} +L_{YM} $$
where now $á$ is covariantized with respect to just the worldsheet metric, we have written the combined U(2) connection explicitly as $g_{Œ'}{}^{º'}$ (but it could be included in the covariant derivative), and the Lagrange multiplier $g^{AB}$ (which is not a U(1) singlet because it multiplies $Ðñ{}^2$, and not $ñÐñ$ like the other terms) enforces the generalized Klein-Gordon equation, which was absent in the Penrose case because it was already ``on-shell".  In particular, this action includes the usual $p^2$ term.

We can put this string on shell by solving the $Ðñ{}^2$ constraint as
$$ Ðñ_{Œ'A} = Ðñ_{Œ'} Ðñ_A $$
in obvious analogy to the Penrose solution in ordinary $x$ space, and then using some of the SU(2) to pick the gauge
$$ Ðñ_{Œ'} = ¶_{Œ'}^{+'} $$
Equivalently, we solve the constraint for $Ðñ_{-'A}$ in terms of $Ðñ_{+'A}$ as
$$ Ðñ_{-'A} ¾ Ðñ_{+'A} $$
and use part of the SL(2) to fix the proportionality constant to 0.  As usual, solving a constraint for a variable allows gauge fixing for the conjugate variable:
$$ constrain¼Ðñ_{-'A} = 0âÜâgauge¼ñ^{-'A} = 0 $$
upon which this action reduces directly to the Berkovits one.  This reduction is analogous to the reduction of the N=4 worldsheet-supersymmetric string to the N=2:  There also reducible constraints are solved to replace a local worldsheet SU(2) (but not U(2)) to a U(1), with irreducible constraints and a lightcone-like formalism.

Ü4. QCD-like superstrings

The explicit expression of a string as a bound state of an underlying particle theory is achieved by random lattice first-quantization of the worldsheet:  The worldsheet (not spacetime) is lattice quantized, but the lattice is irregular to represent the curvature of the worldsheet.  These lattices are then identified with Feynman diagrams of partons that make up the string, and the sum over diagrams with functional integration over the worldsheet metric.  The Feynman rules can be read from the lattice action:  (1) The usual $ü(»X)^2$ term becomes $ü(x_i-x_j)^2$, which produces (because the action appears in an exponential) a propagator $e^{-x^2/2}$ between the two vertices at $x_i$ and $x_j$.  (2) A 2D cosmological term, which gives the worldsheet area, simply counts the number of vertices, so it gives the coupling dependence, with parton coupling equal to the exponential of minus the 2D cosmological constant.  (So vanishing cosmological constant yields unit coupling:  Perturbative relates to nonperturbative when comparing 2D field theory with the parton field theory.)  Since the cosmological term breaks scale invariance, and thus sets the scale, the stringy limit (lattice $£$ continuous worldsheet) is associated with the limit of vanishing cosmological constant.  This parton coupling appears with the ``wrong" sign:  Because of the exponential it always appears as a positive factor, but this corresponds to a negative coupling constant in the parton action.  There can also be other ``geometric" terms in the action, such as an $RÊõ^{-1}R$ term generated by ``hidden matter".  (3) Although only the bosonic string is really understood in this approach, other terms such as Wess-Zumino terms are expected to introduce other contributions, such as derivatives at vertices.

A way to avoid Gaussian propagators, yet still have a Feynman diagram appear as an exponential, is to introduce Schwinger parameters.  Only in D=4 does this work the same way in $x$ space as $p$ space (``T duality"), and only for massless fields, since only then is the $x$-space propagator $1/x^2$.  (Of course, for unphysical Gaussian propagators this invariance under Fourier transformation holds in all D.)  Thus writing
$$ {1\over p^2} = Ç_0^¥ d ¼e^{- p^2}ââorââ
	{1\over x^2} = Ç_0^¥ d÷ ¼e^{-÷ x^2} $$
we are led in the continuum limit to a string action with a ``second metric":  In first-order form, the $X$ part of the action (responsible for the propagator, or at least its denominator) becomes
$$ L ¾ P^à É á_à X +  _{àà}üP^à É P^à $$
(summed over $à$).  There is a separate $ _{++}$ and $ _{--}$ corresponding to the separate propagators leaving a vertex.  (Consider, e.g., a regular square lattice.)  Since $ $ must be diagonal (it gives propagators as squares of $P$'s), it must appear with this 2D Lorentz structure.  The geometric interpretation is that each vertex is 4-point, with propagators in worldsheet-lightlike directions:  For the bosonic string this leads to 4D wrong-sign $Ä^4$ theory, which is asymptotically free.  We have written the second metric $ $ explicitly, while the usual metric appears in the covariant derivative $á$ and the other terms in the action.

Among the twistor superstring's pecularities, there are three that stand out:\\  (1) It describes particles, not strings (no excited states).  (2) Also, it carries only one type of derivative; in the conformal gauge, only $»/»Ðz$ and not $»/»z$.  (3) Furthermore, it has no $Œ'$, since it is scale invariant.  All these suggest that it is a tensionless limit $Œ'£¥$ of a (somewhat) more conventional string.  But it must be an unusual tensionless limit, since $Ðz=§^-$ is a lightlike coordinate:  In the usual string, in first-order form, we take this limit by first rescaling the zweibein (and $å{-g}$) and the $P$'s to obtain
$$ L ¾ P^0 É á_0 X +P^1 É á_1 X -ü(P^0)^2 +Œ'^2 ü(P^1)^2 $$
Taking the $Œ'£¥$ limit then yields a $¶$ function in $P^1$, leaving
$$ L £ P^0 É á_0 X -ü(P^0)^2 $$
describing disjoint particles.  (Alternatively, we could write it in second-order form, and see the term $Œ'^{-2}(á_1 X)^2$ vanish.)  In a Hamiltonian analysis, the $X'ÉP$ constraint remains, but $P^2+Œ'^{-2}X'^2£P^2$, so there is no ``harmonic oscillator potential" joining neighboring pieces of the string.  But there is no way to obtain a residual lightlike derivative:  $P^+$ and $P^-$ appear only in the combination $P^+ÉP^-$.

On the other hand, in the QCD-like string it is $P^+$ and $P^-$ that appear as squares, so we can rescale
$$ L ¾ P^+ É á_+ X +P^- É á_- X +Œ'^2  _{++}ü(P^+)^2 + _{--}ü(P^-)^2 $$
and taking the tensionless limit
$$ L £ P^- É á_- X + _{--}ü(P^-)^2 $$
(Actually this string is conformally invariant, at least classically, so there is no physical $Œ'$, and this parameter merely defines the limit.  However, confinement is expected to generate vacuum values for the $ $'s, giving a physical $Œ'$.)

A more geometric interpretation of the tensionless limit that relates directly to the random lattice is that this limit is the free limit of the parton theory.  This corresponds to the limit where the worldsheet cosmological constant goes to infinity, inducing a factor $¶(å{-g})$.  In other words, the worldsheet area goes to 0, so it becomes a line (or lines).  That makes the worldsheet metric degenerate:  For the usual string, we set $e_1{}^m£0$, killing the $á_1$ term (and making $P^1$ trivial), while for the QCD-like string we set $e_+{}^m£0$, killing the $á_+$ term.

This analysis immediately allows us to derive the ADHM twistor superstring as a tensionless limit from a QCD-like superstring:  The action is
$$ L ¾ (á_à ñ^{Œ'A})Ðñ^à_{Œ'A} 
	+g_{àà}^{AB}Ðñ^{àŒ'}\negthinspace\negthinspace{}_A Ðñ^à_{Œ'B} 
	+g_{àŒ'}{}^{º'} ñ^{Œ'A}Ðñ^à_{º'A} +L_{YM} $$
where $L_{YM}$ also includes $à$ terms.  There is an overall $1/Œ'$, which can be scaled into the ``$+$" terms to eliminate them in the tensionless limit.  On the other hand, if we quantize this action on a random lattice, we obtain partons described by the 1D (worldline) analog of this action, and also allows a Penrose twistor description by solving constraints, which should describe N=4 super Yang-Mills alone, in a way very similar to Berkovits' action.

Ü5. Conclusions

We have introduced two new actions related to the twistor superstring:  (1) One describes a QCD-like superstring, whose partons are themselves N=4 super Yang-Mills.  (2) The other describes the tensionless limit of this QCD-like superstring.  The tensionless limit essentially decouples a parton from the string, so this describes N=4 super Yang-Mills alone.

By solving constraints, the latter action becomes Berkovits' action.  However, in contrast to Berkovits' (or Witten's) action, which contains only 3 on-shell components of external momenta (in the form of a twistor), both our actions include all 4 components of $x$ and $p$.  They use ADHM twistors instead of Penrose twistors (on HP(1|2) instead of CP(3|4)), which are general enough to go off shell, and also describe nonperturbative effects such as instantons.  But they are still twistors in the sense that they transform in the defining representation of the N=4 superconformal group SU(2,2|4) (before gauge fixing the isotropy group U(2)).  As a result, they retain the worldsheet U(1) instantons necessary for the helicity expansion about the selfdual theory.

We have only shown the basic relations between the actions.  Many features remain to be made explicit:  (1) The ADHM action should allow the derivation of Feynman rules in ordinary momentum space, thus allowing a proof of equivalence to the twistor rules of Roiban, Spradlin, and Volovich.  This would be an alternative stringy way to derive Yang-Mills S-matrices.  (2) The possibility of loops in twistor language can be investigated.  (3) A more natural way of incorporating instantons into perturbation theory might be found.  (4) First-quantization of the QCD-like superstring can be examined.  (5) The ADHM twistor actions contain reducible first-class constraints, but avoid the usual second-class constraints of covariant spacetime-supersymmetric actions.  This might make conventional quantization easier.

ÜAcknowledgments

I thank Nathan Berkovits, Riccardo Ricci, and Daniel Robles-Llana for discussions.
This work was supported in part by the National Science Foundation Grant No.¼PHY-0098527.

\refs

£1 M. Ademollo, L. Brink, A. D'Adda, R. D'Auria, E. Napolitano, S. Sciuto,
	E. Del Giudice, P. Di Vecchia, S. Ferrara, F. Gliozzi, R. Musto,
	R. Pettorino, and J.H. Schwarz, \NP 111 (1976) 77.
£2 M. Ademollo, L. Brink, A. D'Adda, R. D'Auria, E. Napolitano, S. Sciuto,
	E. Del Giudice, P. Di Vecchia, S. Ferrara, F. Gliozzi, R. Musto, and
	R. Pettorino, \NP 114 (1976) 297, \PL 62B (1976) 105;\\
	W. Siegel, \xxxlink{hep-th/9204005}, \PR 69 (1992) 1493. 
£3 V.P. Nair, \PL 214B (1988) 215.
£4 E. Witten, \xxxlink{hep-th/0312171}, \xxxlink{hep-th/0403199}.
£5 R. Roiban, M. Spradlin, and A. Volovich, \xxxlink{hep-th/0402016},
	{\refit JHEP} {\refbf 0404} (2004) 012, \xxxlink{hep-th/0403190};\\
	R. Roiban and A. Volovich, \xxxlink{hep-th/0402121}.
£6 N. Berkovits, \xxxlink{hep-th/0402045};\\
	N. Berkovits and L. Motl, \xxxlink{hep-th/0403187}.
£7 H. Ooguri and C. Vafa, ÓMod. Phys. Lett.Õ É{A5} (1990) 1389,
	\NP 361 (1991) 469, {\refbf 367} (1991) 83;\\
	N. Marcus, \xxxlink{hep-th/9207024}, \NP 387 (1992) 263.
£8 W. Siegel, \xxxlink{hep-th/9601002}, ÓInt. J. Mod. Phys. AÕ É13 (1998)
	381.
£9 H.B. Nielsen and P. Olesen, \PL 32B (1970) 203;\\
	D.B. Fairlie and H.B. Nielsen, \NP 20 (1970) 637;\\
	B. Sakita and M.A. Virasoro, \PR 24 (1970) 1146;\\
	F. David, \NP 257 [FS14] (1985) 543;\\
	V.A. Kazakov, I.K. Kostov, and A.A. Migdal, \PL 157B (1985) 295;\\
	J. Ambj\o rn, B. Durhuus, and J. Fršhlich, \NP 257 (1985) 433;\\
	M.R. Douglas and S.H. Shenker, \NP 335 (1990) 635;\\
	D.J. Gross and A.A. Migdal, \PR 64 (1990) 127;\\
	E. Br«ezin and V.A. Kazakov, \PL 236B (1990) 144.
£10 P. De Causmaecker, R. Gastmans, W. Troost, and T.T. Wu, 
	\NP 206 (1982) 53;\\
	F.A. Berends, R. Kleiss, P. De Causmaecker, R. Gastmans, W. Troost,
	and T.T. Wu, \NP 206 (1982) 61;\\
	Z. Xu, D.-H. Zhang, and L. Chang, \NP 291 (1987) 392;\\
	J.F. Gunion and Z. Kunszt, \PL 161 (1985) 333;\\
	R. Kleiss and W.J. Sterling, \NP 262 (1985) 235.
£11 S.J. Parke and T. Taylor, \NP 269 (1986) 410, \PR 56 (1986) 2459;\\
	F.A. Berends and W.T. Giele, \NP 306 (1988) 759.
£12 W. Siegel, \xxxlink{hep-th/9205075}, \PRD 46 (1992) R3235;
	\xxxlink{hep-th/9207043}, \PRD 47 (1993) 2504.
£13 G. Chalmers and W. Siegel, \xxxlink{hep-ph/9801220},
	\PRD 59 (1999) 045013.
£14 R. Brooks (May, 1993), unpublished;\\
	G. Chalmers and W. Siegel, \xxxlink{hep-th/9606061}, \PRD 54 (1996)
	7628;\\
	N. Berkovits and W. Siegel, \xxxlink{hep-th/9703154}, \NP 505 (1997)
	139.
£15 R. Penrose, ÓJ. Math. Phys.Õ É8 (1967) 345,
	ÓInt. J. Theor. Phys.Õ É1 (1968) 61;\\
	M.A.H. MacCallum and R. Penrose, ÓPhys. Rep.Õ É6 (1973) 241.
£16 A. Ferber, \NP 132 (1978) 55.
£17 M.F. Atiyah and R.S. Ward, ÓComm. Math. Phys.Õ É55 (1977) 117;\\
	M.F. Atiyah, V.G. Drinfel'd, N.J. Hitchin, and Yu.I. Manin, \PL 65A (1978)
	185;\\
	E. Corrigan, D. Fairlie, P. Goddard, and S. Templeton, \NP 140 (1978)
	31;\\
	N.H. Christ, E.J. Weinberg, and N.K. Stanton, \PRD 18 (1978) 2013;\\
	M.F. Atiyah, ÓGeometry of Yang-Mills fieldsÕ (Scuola Normale
	Superiore, Pisa, 1979);\\
	V.E. Korepin and S.L. Shatashvili, ÓMath. USSR IzvestiyaÕ É24 (1985)
	307.
£18 W. Siegel, \xxxlink{hep-th/9412011}, \PRD 52 (1995) 1042.
£19 A. Schild, \PRD 16 (1977) 1722;\\
	A. Karlhede and U. Lindstr¬om, 
	{\refit Class. Quant. Grav.} {\refbf 3} (1986) L73;\\ 
 	F. Lizzi, B. Rai, G. Sparano, and A. Srivastava, \PL 182B (1986) 326;\\
 	A.A. Zheltukhin, {\refit Sov. J. Nucl. Phys.} {\refbf 48} (1988) 375;\\
 	A. Barcelos-Neto and M. Ruiz-Altaba, \PL 228B (1989) 193;\\
	J. Gamboa, C. Ramirez, and M. Ruiz-Altaba, \NP 338 (1990) 143;\\
	U. Lindstr¬om, B. Sundborg, and G.Theodoridis, \PL 253B (1991) 319,
	{\refbf 258B} (1991) 331;\\
 	M. Ro×cek and U. Lindstr¬om, \PL 271B (1991) 79;\\
	J. Isberg, U. Lindstr¬om, and B. Sundborg, \xxxlink{hep-th/9207005}, 
	\PL 293B (1992) 321;\\ 
	 J. Isberg, U. Lindstr¬om, B. Sundborg, and G. Theodoridis,
	 \xxxlink{hep-th/9307108}, \NP 411 (1994) 122. 
£20 J. Schwinger, On angular momentum, ÓQuantum theory of angular
	momentum: a collection of reprints and original papersÕ, eds. L.C.
	Biedenharn and H. Van Dam (Academic, 1965) p. 229.
£21 K. Bardak'ci and M.B. Halpern, \PRD 3 (1971) 2493.
£22 N. Berkovits, \xxxlink{hep-th/9208035}, \NP 395 (1993) 77;\\
	J. Bischoff and O. Lechtenfeld, \xxxlink{hep-th/9612218}, 
	\PL 390B (1997) 153;\\
	O. Lechtenfeld and W. Siegel, \xxxlink{hep-th/9704076}, \PL 405B (1997) 49.
£23 S. Mandelstam, \NP 213 (1983) 149;\\
	L. Brink, O. Lindgren, and B.E.W. Nilsson, \NP 212 (1983) 401.
£24 M.B. Green and J.H. Schwarz, \NP 243 (1984) 475.

\bye